\documentclass[useAMS,usenatbib]{mn2e}
 
\usepackage{graphicx}

\title[Probing the USco IMF in the planetary-mass regime]
{Probing the Upper Scorpius mass function in the planetary-mass regime
\thanks{Based on observations collected with the ESO VISTA telescope
under programme number 089-C.0102(ABC).}
\author[N. Lodieu]{N. Lodieu$^{1,2}$\thanks{E-mail: nlodieu@iac.es},
P.\ D.\ Dobbie$^{3}$, N. J. G. Cross$^{4}$, N.\ C.\ Hambly $^{4}$,
M.\ A.\ Read$^{4}$,
 \newauthor
R.\ P.\ Blake$^{4}$, D.\ J.\ E.\ Floyd$^{5}$ \\
$^{1}$ Instituto de Astrof\'isica de Canarias (IAC), C/ V\'ia L\'actea s/n, 
E-38200 La Laguna, Tenerife, Spain \\
$^{2}$ Departamento de Astrof\'isica, Universidad de La Laguna (ULL),
E-38206 La Laguna, Tenerife, Spain \\
$^{3}$ School of Mathematics \& Physics, University of Tasmania, Hobart, TAS, 7001, Australia \\
$^{4}$ Scottish Universities Physics Alliance (SUPA), Institute for Astronomy, School of Physics and Astronomy, \\
University of Edinburgh, Royal Observatory, Blackford Hill, Edinburgh EH9 3HJ, UK \\
$^{5}$ School of Physics, Monash University, Clayton, Victoria 3800, Australia}
}
\begin{document}

\date{Accepted \today. Received \today; in original form \today}

\pagerange{\pageref{firstpage}--\pageref{lastpage}} \pubyear{2005}

\maketitle

\label{firstpage}

%
%
\begin{abstract}
We present the results of a deep $ZYJ$ near-infrared survey of 13.5 square 
degrees in the Upper Scorpius (USco) OB association. We photometrically selected  
$\sim$100 cluster member candidates with masses in the range 30--5 Jupiters,
according to state-of-the-art evolutionary models. We identified 67
$ZYJ$ candidates as bona-fide members, based on complementary photometry 
and astrometry. We also extracted five candidates detected with VISTA 
at $YJ$ only. One is excluded using deep optical $z$-band imaging, while 
two are likely non-members, and three remain as potential members. We conclude 
that the USco mass function is more likely decreasing in the planetary-mass 
regime (although a flat mass function cannot yet be discarded), consistent 
with surveys in other regions.
\end{abstract}


%
%
\section{Introduction}
\label{USco_VISTA:intro}

The quest for young low-mass objects in the planetary-mass regime aims at 
addressing two fundamental questions in our understanding of star formation: 
what is the lowest mass fragment that star formation processes can form? and 
what is the shape of the initial mass function \citep[IMF;][]{salpeter55,miller79,scalo86} 
below the deuterium-burning limit? Finding planetary-mass objects can help 
distinguishing among the different star formation models proposed during the 
past decades. The earliest theoretical prediction by \citet{kumar69} suggested 
a minimum mass of one Jupiter mass (M$_{\rm Jup}$) whereas 3D opacity-limited 
hierarchical fragmentation yielded a range between 7 and 10 M$_{\rm Jup}$
\citep{low76,rees76,silk77a,boss88}. 
More recent calculations place that limit around (or below) 3--5 M$_{\rm Jup}$
\citep{boyd05,whitworth06,forgan11,rogers12}.

Several surveys have been carried out over the past years to identify bona-fide
L and T dwarfs in young star-forming regions to probe the planetary-mass regime
and investigate the shape of the substellar mass function. Old field L dwarfs
with ages greater than 1 Gyr have typically effective temperatures in the
2200--1400\,K range and (model-dependent) masses around 85--50 M$_{\rm Jup}$
\citep{basri00,leggett00,kirkpatrick05}, whereas old T dwarfs have temperatures 
below 1400\,K \citep{burgasser06a} and masses below $\sim$50 M$_{\rm Jup}$ 
\citep{kirkpatrick05}. By comparison, 10 Myr-old L and T dwarfs would have
masses in the $\sim$20--4 M$_{\rm Jup}$ and $\leq$4 M$_{\rm Jup}$, respectively
\citep{kirkpatrick05}. The best studied region for young
L and T dwarfs is $\sigma$\,Orionis, where deep photometric surveys 
complemented by optical and near-infrared spectroscopy have been conducted
\citep{zapatero00,barrado01c,martin01a,caballero07d,bihain09,penya11a,penya12a}.
The mass function in this region keeps rising in a linear scale 
\citep[dN/dM$\propto$M$^{-{\alpha}}$;][]{kroupa02} from early-type stars 
(19 solar masses) all the way down to the planetary-mass regime 
($\leq$13 M$_{\rm Jup}$) with a possible 
turn-over below 5 M$_{\rm Jup}$ \citep{penya12a}. In NGC\,1333, \citet{scholz12c} 
estimated that the number of 5--15 M$_{\rm Jup}$ members should be less than ten, 
implying a decreasing mass function in the planetary-mass regime and a possible cut-off
below 6 M$_{\rm Jup}$. \citet{alves12} confirmed spectroscopically the first
L dwarfs in the $\rho$\,Ophiuchus molecular cloud with masses between 
10 and 4 M$_{\rm Jup}$, resulting in a mass function in agreement with 
other regions \citep{bastian10} and the field \citep{chabrier03,chabrier05a}. 
Moreover, \citet{spezzi12b} reported two potential T--type candidates in the core 
of the Serpens cloud with masses of a few Jupiter masses if true members. 
Similarly \citet{burgess09} identified one mid-T--type candidate in IC\,348 
with a mass between 5 and 1 M$_{\rm Jup}$ but spectroscopy is mandatory 
to confirm their nature. This T-type candidate is consistent with the extrapolation 
of the substellar IMF in IC\,348 \citep{alves13a}.
A few other L and T dwarf candidates have been announced in older clusters
\citep[e.g.][]{casewell07,bouvier08a,hogan08,boudreault13} but state-of-the-art models anticipate 
masses in the midst of the substellar regime \citep[30--50 M$_{\rm Jup}$;][]{baraffe98,chabrier00c}
because of their older ages \citep[125 Myr for the Pleiades and $\sim$600 Myr for 
Praesepe and the Hyades;][]{stauffer98,perryman98,fossati08}.

Upper Scorpius (hereafter USco) is part of the nearest OB association to the
Sun, Scorpius Centaurus. The combination of its distance 
\citep[145 pc;][]{deBruijne97}, young age
\citep[5 or 10 Myr depending on the authors;][]{preibisch99,pecaut12,song12},
and proper motion \citep[mean value of $-$25 mas/yr;][]{deBruijne97,deZeeuw99},
makes it an ideal place to look for brown dwarfs and isolated planetary-mass
objects to test the theory of the fragmentation limit 
\citep{kumar69,low76,rees76}. The bright end of the USco population has been 
examined in X-rays \citep{walter94,kunkel99,preibisch98}, astrometrically 
\citep{deBruijne97,deZeeuw99}, and spectroscopically \citep{preibisch02}. 
The low-mass and substellar population has been investigated over the past 
decade in great details with the advent of modern detectors permitting wide 
and/or deep surveys
\citep{ardila00,martin04,slesnick06,lodieu06,lodieu07a,slesnick08,dawson11,lodieu11c,dawson12,lodieu13c}.
More recently, we published five potential T--type candidates \citep{lodieu11c},
discovery not supported by subsequent optical and near-infrared imaging
\citep{lodieu13b}.

In this paper, we present a deep $ZYJ$ ($J$\,=\,20.5 mag 100\% completeness) 
and wide (13.5 square degrees) near-infrared survey of the central region of 
the USco association conducted with the Visible and Infrared Survey Telescope 
for Astronomy \citep[VISTA;][]{emerson01,emerson04} located in Paranal, Chile.
In Section \ref{USco_VISTA:survey} we describe the observations, data
reduction, and image stacking procedure.
In Section \ref{USco_VISTA:other_survey} we describe complementary datasets
from the United Kingdom InfraRed Telescope (UKIRT) Infrared Deep Sky 
Survey \citep[UKIDSS;][]{lawrence07} Galactic Clusters Survey (GCS) and 
optical imaging from the 6.5-m Magellan telescope. 
In Section \ref{USco_VISTA:select} we detail the photometric 
selection of member candidates in the USco association.
In Section \ref{USco_VISTA:MF} we discuss the shape of the mass
function in the planetary-mass regime, comparing the numbers of
candidates with our previous determination from the UKIDSS GCS\@.

%
%
\section{The deep $ZYJ$ survey}
\label{USco_VISTA:survey}
\subsection{Near-infrared imaging}
\label{USco_VISTA:survey_obs}

\begin{figure}
  \centering
  \includegraphics[width=\linewidth, angle=0]{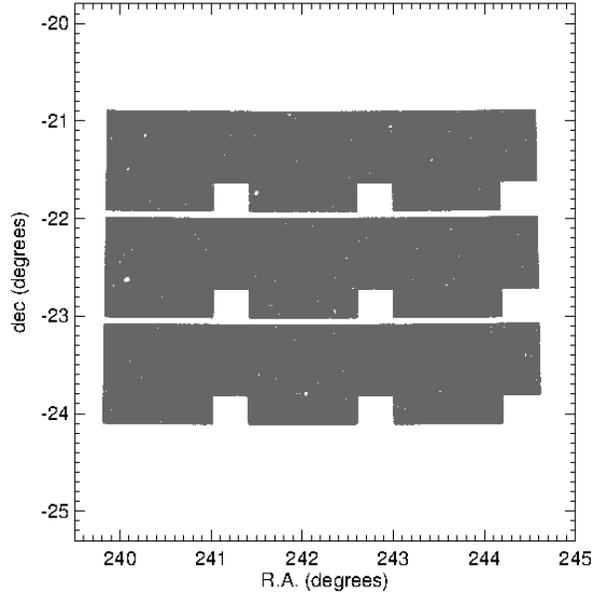}
  \caption{Coverage of the deep VISTA survey in the USco association:
13.5 square degrees imaged in $ZYJ$.}
  \label{fig_dT_PM:cover}
\end{figure}

We performed a near-infrared survey of 13.5 square degrees in USco 
(Fig.\ \ref{fig_dT_PM:cover}) with the VISTA InfraRed CAMera
\citep[VIRCAM;][]{dalton06} mounted on VISTA telescope 
\citep{emerson01,emerson04} in Paranal, Chile. We choose this
part of USco because a subset of this region was imaged as part of the
UKIDSS GCS Science verification \citep{lodieu07a} and contains many
low-mass brown dwarf members confirmed spectroscopically \citep{lodieu11a}. 
VIRCAM contains 67 million pixels of mean size 0.339 arcsec, offering an
instantaneous field-of-view of 0.6 square degrees. Six ``paw-print''
observations are mosaicked together to produce a tile with an area of 1.65 
square degrees. We imaged USco in the $ZYJ$ filters in service mode between
April and May 2012 (see logs of observations in 
Table \ref{tab_USco_VISTA:log_obs}). In this paper 
we use only observing blocks completed as part of our ESO 
(European Southern Observatory) program 089.C-102(ABC) that complied with our 
original request (grade A or B according to ESO observing logs): clear sky, 
grey time with the moon further away than 30 degrees from our target to avoid 
reflection on the telescope, seeing better than 1.2 arcsec, and airmass less 
than 1.5 (Table \ref{tab_USco_VISTA:log_obs}).

We employed a similar strategy for all three passbands. We set the detector 
on-source integrations (DIT) to 50 sec, 60 sec, and 24 sec in $Z$, $Y$,
and $J$, respectively, with six pawprint positions and five jitter positions
to create a sky frame (no microstepping). We repeated the DIT twice in the 
case of $Z$ and also repeated the full observing blocks. The total exposure 
times amount for 100 min, 30 min, and 12 min in $Z$, $Y$, and $J$, respectively.

%
%
%

%
%
\begin{figure}
  \centering
  \includegraphics[width=\linewidth, angle=0]{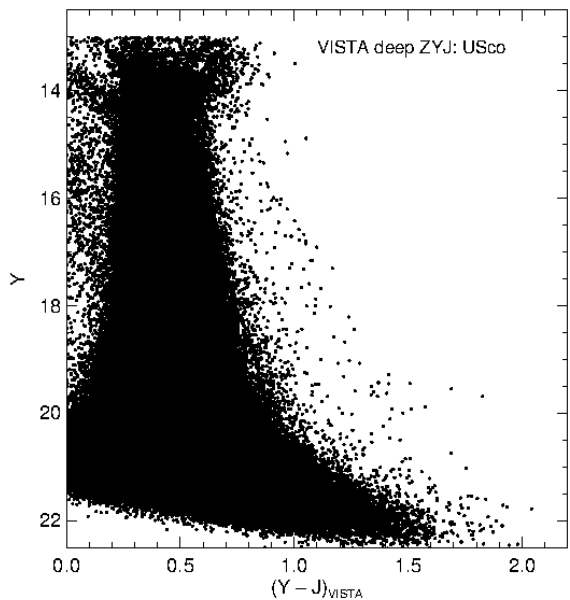}
  \caption{($Y-J$,$Y$) colour-magnitude diagram for all sources detected in 
the VISTA deep survey.}
  \label{fig_dT_PM:YJY_alone}
\end{figure}
%

%
%
\begin{table*}
 \centering
 \caption[]{Log of the VISTA/VIRCAM observations. We list the central 
coordinates of each tile, followed by the dates, averaged seeing, mean 
ellipticities (Ell), and photometric zero-points (ZP) for each tile and
each filter.}
{\scriptsize
 \begin{tabular}{@{\hspace{0mm}}c | c c | c c c c | c c c c | c c c c@{\hspace{0mm}}}
 \hline
 \hline
Tile & R.A.    &     Dec       &   \multicolumn{4}{c}{Z} & \multicolumn{4}{c}{Y} & \multicolumn{4}{c}{J} \cr
 \hline
      &             &                     &   Date   & Seeing & Ell & ZP & Date & Seeing & Ell & ZP & Date & Seeing & Ell & ZP   \cr
\hline
      & hh:mm:ss.ss & ${^\circ}$:$'$:$''$ & yymmdd & arcsec &  & mag & yymmdd & arcsec &  & mag & yymmdd & arcsec &  & mag   \cr
 \hline
1\_1\_1 & 16:02:05.18 & $-$23:41:04.6 & 120428 & 0.974 &  0.040 & 23.761 & 120415 & 1.302 & 0.029 & 23.471 & 120420 & 0.955 & 0.079 & 23.752 \cr
1\_1\_2 & 16:08:27.20 & $-$23:41:04.6 & 120517 & 0.989 &  0.052 & 23.762 & 120419 & 1.000 & 0.042 & 23.460 & 120420 & 0.944 & 0.048 & 23.738 \cr
1\_1\_3 & 16:14:43.50 & $-$23:41:04.6 & 120516 & 0.905 &  0.052 & 23.754 & 120417 & 0.821 & 0.061 & 23.466 & 120428 & 0.883 & 0.064 & 23.753 \cr
1\_2\_1 & 16:02:05.18 & $-$22:35:33.0 & 120430 & 0.885 &  0.053 & 23.770 & 120417 & 0.960 & 0.042 & 23.432 & 120421 & 0.703 & 0.060 & 23.766 \cr
1\_2\_2 & 16:08:27.20 & $-$22:35:33.0 & 120430 & 0.836 &  0.064 & 23.774 & 120416 & 0.769 & 0.058 & 23.478 & 120430 & 0.899 & 0.048 & 23.757 \cr
1\_2\_3 & 16:14:46.40 & $-$22:35:33.0 & 120516 & 0.867 &  0.054 & 23.754 & 120419 & 0.775 & 0.054 & 23.463 & 120428 & 0.951 & 0.049 & 23.738 \cr
1\_3\_1 & 16:02:05.18 & $-$21:30:01.2 & 120517 & 0.935 &  0.050 & 23.766 & 120417 & 0.777 & 0.059 & 23.478 & 120422 & 0.778 & 0.056 & 23.746 \cr
1\_3\_2 & 16:08:27.20 & $-$21:30:24.8 & 120516 & 1.012 &  0.044 & 23.777 & 120416 & 0.769 & 0.055 & 23.479 & 120422 & 0.781 & 0.061 & 23.744 \cr
1\_3\_3 & 16:14:49.66 & $-$21:29:53.3 & 120516 & 0.971 &  0.042 & 23.768 & 120418 & 0.832 & 0.053 & 23.473 & 120422 & 0.762 & 0.056 & 23.736 \cr
  \hline
 \label{tab_USco_VISTA:log_obs}
 \end{tabular}
}
\end{table*}
\subsection{Data reduction}
\label{USco_VISTA:survey_DR}
The data reduction and processing was done as part of the VISTA Data Flow System
\citep[VDFS;][]{emerson04}, which involves initial quality control at
Garching by ESO, observing block (OB) processing and monthly photometric
calibration at Cambridge by the Cambridge Astronomical Survey Unit (CASU)
\footnote{See http://casu.ast.cam.ac.uk/surveys-projects/vista/technical/ 
for the latest details} and multi-night processing, matching to external data
sets  and archiving in Edinburgh by Wide-Field Astronomy Unit \citep[WFAU;][]{cross12}.

The initial data reduction of the VISTA images involves the following processes\footnote{More
details at http://apm49.ast.cam.ac.uk/surveys-projects/vista/technical/data-processing}:
\begin{itemize}
   \item Reset correction - similar to bias removal
   \item Dark correction - removal of dark current, using exposures taken through a dark filter
   \item Linearity correction - take out non-linearities in the detector response
   \item Flat field correction - remove small scale QE variations using twilight flats
   \item Sky background correction - remove 2D sky background
   \item Destripe - remove stripe pattern from readout channels. Use detector symmetry 
   \item Jitter stacking - combine individual, but offset, exposures to remove defects
   \item Catalogue generation - extractor finds sources that have more than 4
   connected pixels at a threshold of $1.5\sigma$ above the sky noise.
   \item Astrometric and photometric calibration - Uses the 2MASS all sky
   catalogue to get an astrometric and photometric calibration. Enhancements,
   such as zeropoint offsets between detectors and the scattered light
   (illumination) correction are found later, using a month of data. VISTA data
   has not been accurately calibrated to the Vega system, as in the case of
   UKIRT-WFCAM. Instead it was decided to leave the photometry in the VISTA
   system and provide colour equations, see Hodgkin et al.\ (2013, in prep).
   \item Tile generation - 6 pawprints are mosaicked together to form a 1.5 sq.
   degree tile. The catalogue generation is done from a different tile, created
   after filtering each pawprint with a nebulosity filter to remove structure on
   scales $>30\arcsec$, to produce smooth backgrounds. Additional steps of
   grouting, to remove effects of mosaicing together pawprints with different
   PSFs and reclassification are necessary to produce the best tile catalogues.
\end{itemize} 

All of the individual OBs in the VISTA data are processed in this way. Automated
quality control was run to deprecate poor quality frames. More than $75\%$ of
all science frames in this programme passed with over $75\%$ of the deprecated 
frames coming from OBs which were incomplete or otherwise were set to be
reobserved (ESOGRADE C/D/R). At WFAU, OB tiles are grouped into the individual
programmes and then processed programme by  programme \citep{cross12} matching
data across filters and epochs. The main steps are:

\begin{itemize}
  \item Create tile-pawprint tables linking tile detections to pawprint
  detections.
  \item Create deep stacks and tiles. This was necessary for some $Z$-band data
  which was split over two OBs. 
  \item Band merge deepest data in each band to produce \verb+n089c102Source+
  table, which is seamed to create a list of unique objects
  \item Create neighbour tables between \verb+n089c102Source+ and external
  surveys: UKIDSS DR10 GCS \citep{lawrence07}, 2MASS point source catalogue
  \citep{cutri03,skrutskie06}, and SuperCOSMOS Science Archive \citep{hambly01a}.
  \item Create best-match and variability tables, so that light-curves can be
  generated for sources. This was done for the $Z$-band with the 2-epochs, with
  some sources (in overlap regions) having up to eight observations.
\end{itemize}

\subsection{Image stacking and tiling}
\label{USco_VISTA:survey_stacking}

The creation of deep pawprint stacks and deep tiles from data across many OBs
uses the same software (developed by CASU
\footnote{http://apm49.ast.cam.ac.uk/surveys-projects/software-release}) as was
used in the OB stacking and tile creation, except that is bound into the VISTA Science 
archive software, which selects all the data in the same pointing and filter
that passes the quality control, rather than using data with the same OB and
group identifiers. The processing of deep stacks and tiles is described in
detail in \cite{cross12}. Deep tiles are not made by stacking OB tiles, but are
created by mosaicing 6 deep pawprint stacks which are in turn created by
stacking OB pawprint stacks. This processing sequence gives the best image quality. 
Photometric calibration of deep pawprint stacks is done by comparing to the 
component OB stacks; these having been observed in the same filter avoids any
colour selection or colour terms that are necessary when comparing with 2MASS. 
Similarly deep tiles are calibrated by comparing to the component deep pawprints.

The ($Y-J$,$Y$) colour-magnitude diagram for all high-quality point sources 
after the stacking process is shown in Fig.\ \ref{fig_dT_PM:YJY_alone}.
We estimate the depth of our final combined images from the point where the 
linear fit to the histograms of the numbers of sources as a function of 
magnitude diverge. We derive 100\% completeness limits of 22.0 mag, 21.2 mag, 
and 20.5 mag in $Z$, $Y$, and $J$, respectively. The saturation limits are
14.0 mag, 14.6 mag, and 14.3 mag in $Z$, $Y$, and $J$, respectively.

%
%
%
\section{Complementary datasets}
\label{USco_VISTA:other_survey}
\subsection{The UKIDSS GCS as first epoch}
\label{USco_VISTA:other_survey_GCS}

The UKIDSS project, defined in \citet{lawrence07}, consists of several 
sub-projects, including the GCS whose main scientific goal is to study the 
shape and universality of the mass function in the low-mass and substellar
regimes. UKIDSS uses the UKIRT Wide Field 
Camera \citep[WFCAM;][]{casali07}. The photometric system is described in
\citet{hewett06}, and the calibration is described in \citet{hodgkin09}.
The pipeline processing and science archive are described in \citet{irwin04}
and \citet{hambly08}, respectively. In this section, we use the GCS
data to complement our deep $ZYJ$ VISTA survey with $H$ and $K$ photometry
and proper motions from the tenth data release (14 January 2013).
The 5$\sigma$ completeness limits of the UKIDSS GCS are typically
$Z$\,=\,20.2 mag, $Y$\,=\,19.8 mag, $J$\,=\,19.3 mag, $H$\,=\,18.4 mag, and 
$K$\,=\,18.1 mag \citep[e.g.\ Figure 2 of][]{lodieu12a,dye06}, implying that 
our VISTA survey is 1.2--1.8 mag deeper than the GCS depending on the filter.
The proper motions computed as part of the latest GCS data release take
into account the different epochs of observations in all the filters
\citep{collins12}, providing proper motions accurate to 5--10 mas/yr.

\subsection{Magellan/IMACS $z$-band imaging}
\label{USco_VISTA:other_survey_IMACS}

A number of regions of Upper Sco were observed in the $z$-band during 
May 2009 with the Inamori Magellan Areal Camera and Spectrograph (IMACS) 
and the 6.5m Magellan Baade telescope. Total integration times of 90 min 
were built-up through a number of 300s sub-exposures. IMACS is a wide-field 
imaging system (or multiobject spectrograph) in which light from the 
Gregorian secondary of the telescope is fed to one of two cameras, short 
(f/2.5) or long (f/4.3). The short camera, used for these observations, 
is constructed around a mosaic of eight 2048$\times$4096 pixel $E2V$ CCDs 
and covers a 27.5'$\times$27.5' (0.20 arcsec pixels) area of sky per 
pointing. However, it suffers from strong vignetting at distances 
more that 15' from the field center.
 
The data frames were reduced using the Cambridge Astronomical Survey Unit CCD 
reduction toolkit \citep{irwin01} to follow standard steps, namely, subtraction 
of the bias, flat-fielding, astrometric calibration, and co-addition. We 
performed aperture photometry on the reduced images using a circular window 
with a diameter of 1.5$\times$ the full width half maximum of the mean point 
spread function ($\sim$0.6-0.7 arcsec). Finally, we cross-correlated our 
optical photometry with our catalogue of near-infrared data and transformed our 
instrumental magnitudes onto approximately the VISTA $Z$-band natural system. 

%
%
\begin{figure*}
  \centering
  \includegraphics[width=0.49\linewidth, angle=0]{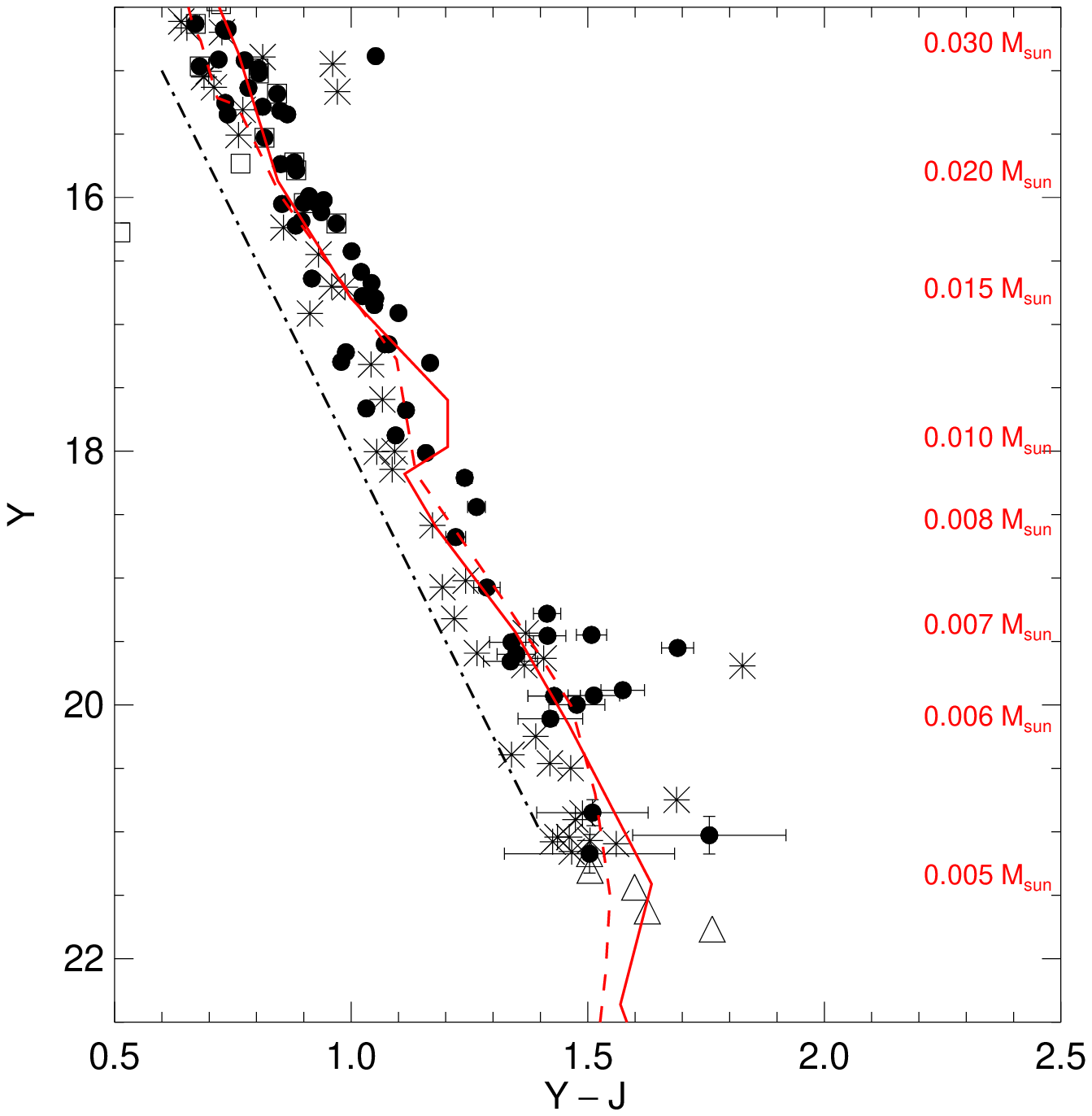}
  \includegraphics[width=0.49\linewidth, angle=0]{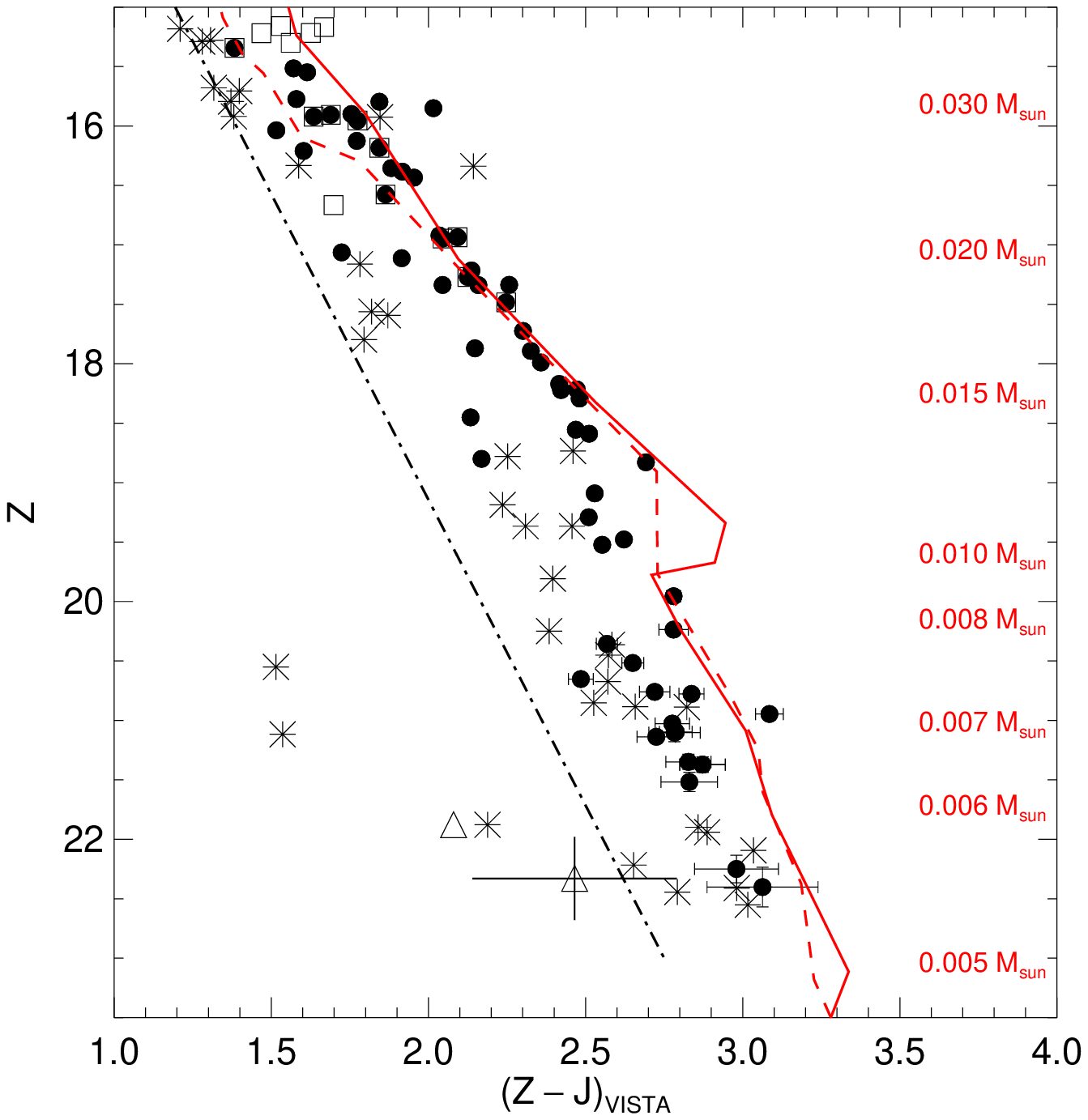}
  \includegraphics[width=0.49\linewidth, angle=0]{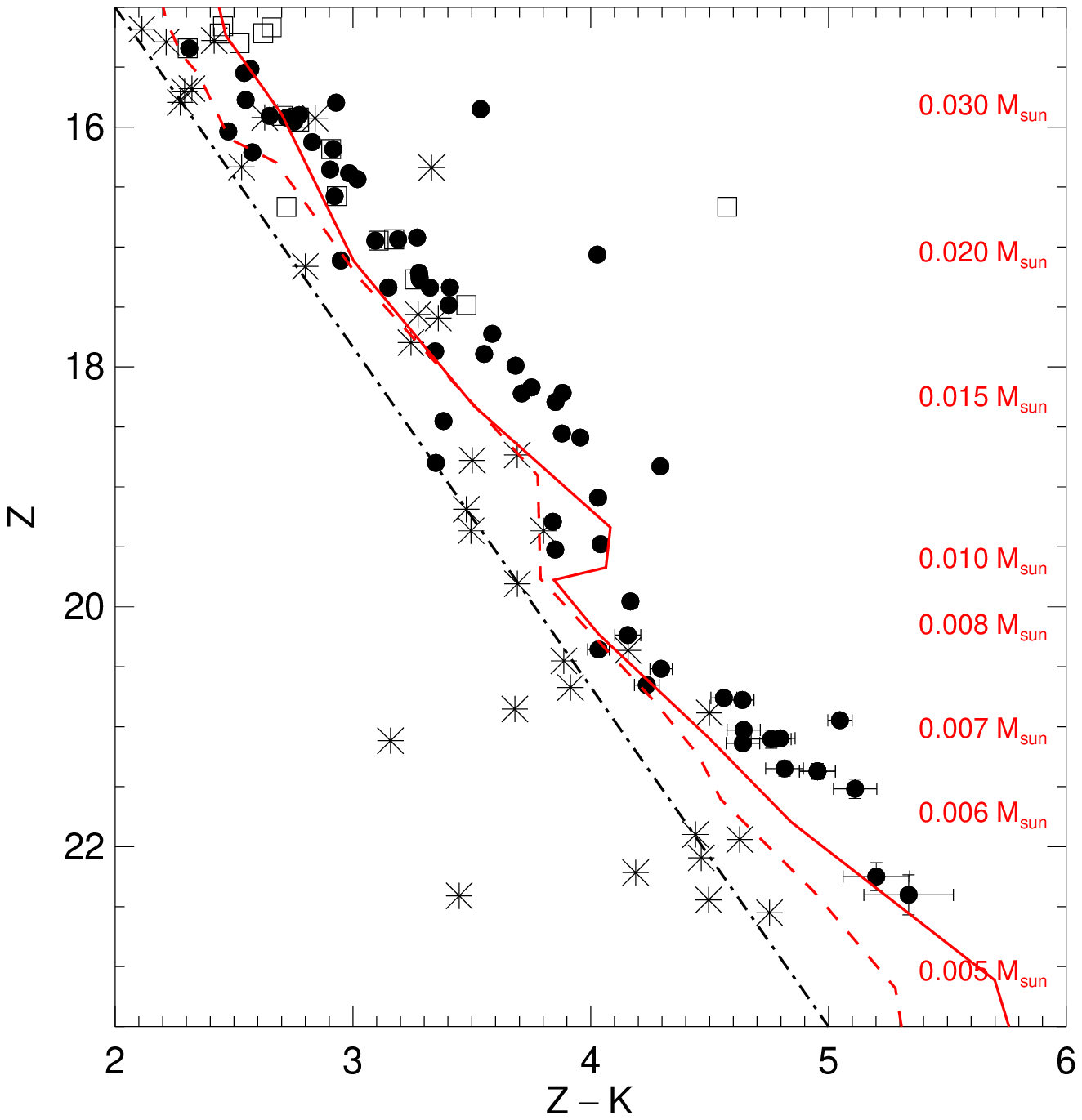}
  \includegraphics[width=0.49\linewidth, angle=0]{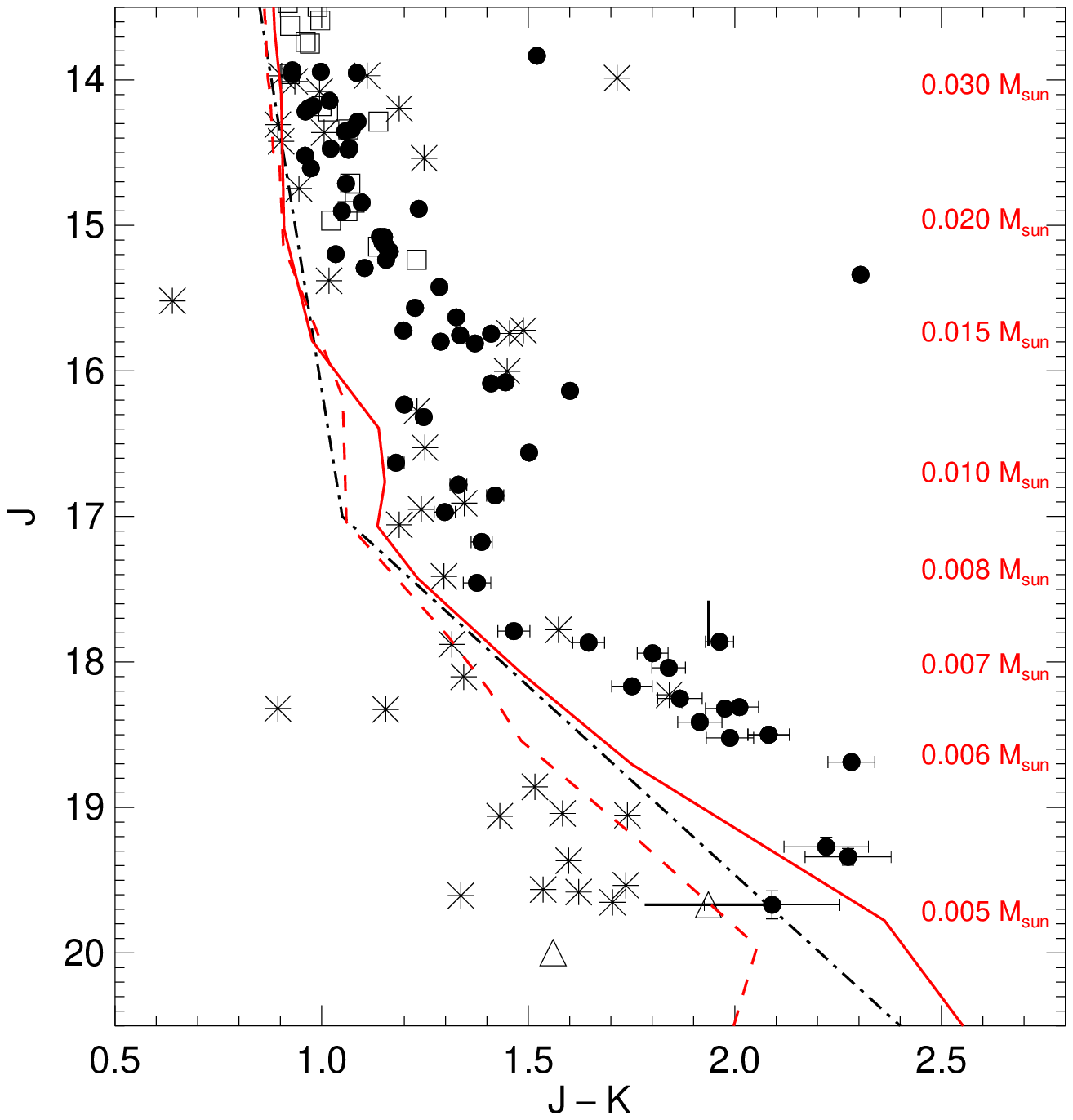}
  \caption{($Y-J$,$Y$), ($Z-J$,$Z$), ($Z-K$,$Z$), and ($J-K$,$J$) 
colour-magnitude diagrams (ordered following the photometric selection
described in this work) for member candidates identified photometrically 
in the VISTA (filled dots). Open squares are previously-confirmed USco 
spectroscopic members \citep{lodieu11a}. Photometric non-members are 
marked with an asterisk. $YJ$-only candidates with $Z$-band drop-outs 
are plotted as open triangles. Uncertainties in magnitudes and colours 
are included only for member candidates: some error bars are not visible 
because they are smaller than the size of the dots. Overplotted are the 
5 and 10 Myr BT-Settl isochrones (red lines) shifted at the distance of 
USco \citep{allard12}. The associated masses are quoted on the right-hand
side, assuming an age of 5 Myr and a distance of 145 pc for USco.}
  \label{fig_dT_PM:CMDs}
\end{figure*}
%

%
%
\section{Photometric selection of member candidates in USco}
\label{USco_VISTA:select}
\subsection{Contents of the VISTA catalogue}
\label{USco_VISTA:select_cat}

We wrote a Structure Query Language sent to the VISTA Science archive 
\citep{cross12} to retrieve all good-quality point sources present in the 
VISTA $ZYJ$ catalogue. We imposed quality criteria set by the {\tt{ppErrBits}}
parameters (less than 256) as well as point source criteria determined by the
{\tt{Class}} and {\tt{ClassStat}} parameters. We considered only sources
fainter than $Z,Y$\,=\,13 mag and $J$\,=\,12 mag although saturation limits
are fainter by at least one magnitude. We also allowed for $Z$-band non
detection to improve our completeness at the faint end.

Our deep VISTA survey overlaps with the $HK$ coverage from the GCS DR10\@.
Our SQL query includes a cross-match of both catalogues using a matching radius 
of 3 arcsec. Hence, every source in the VISTA catalogue is associated with
an object in the GCS catalogue if there is one within this matching radius. 
The final catalogue provides $ZYJHK1K2$ photometry (two epochs in $K$) 
for 1,098,027 point sources\footnote{we plan to make this catalogue available
to the community via Vizier at the Centre de Donn\'ees de Strasbourg}, 
along with their proper motions and associated errors from the GCS DR10\@.

These proper motions and their
associated errors are measured from the multiple epochs available from the GCS, 
calculating a mean epoch and a making linear least-square fit on the errors
of the positions with standard error propagation. A factor for the noise floor
has been added to correct for the fact that the centroid errors are not getting
better with brighter magnitudes but instead reach a plateau. The procedure
is detailed at length in \citet*{collins12}.

\subsection{Deep $ZYJ$ survey}
\label{USco_VISTA:select_ZYJ}

We applied photometric cuts in two colour-magnitude diagrams to select 
member candidates in the 13.5 square degrees surveyed in USco, based on the 
location of known USco members located in the VISTA coverage (open squares 
in Fig.\ \ref{fig_dT_PM:CMDs}), which are confirmed photometrically, 
astrometrically, and spectroscopically \citep{lodieu11a,lodieu13c}.
We recovered 37 known members from our previous work, including 12 fainter 
than $Y$\,=\,14.6 mag \citep[open square in 
Figs \ref{fig_dT_PM:CMDs} \& \ref{fig_dT_PM:CCD};][]{lodieu11a}.

We restrain our photometric selection to sources fainter than $Y$\,=\,14.6 mag
(unsaturated objects) and brighter than the 100\% completeness limits of
our survey in $Y$ and $J$ only. We started our photometric selection in the 
($Y-J$,$Y$) colour-magnitude diagram (Fig.\ \ref{fig_dT_PM:CMDs})
because it offers the greatest sensitivity to substellar members. 
We kept only sources to the right of the lines (dot-dashed lines in 
Fig.\ \ref{fig_dT_PM:CMDs}) defined by:
\begin{itemize}
\item ($Y-J$,$Y$) = (0.6,15.0) to (1.4,21.0)
\item ($Z-J$,$Z$) = (1.0,14.0) to (2.75,23.0)
\end{itemize}

The $YJ$ selection returned 109 candidates, number reduced to 101 when
applying the photometric cuts in the ($Z-J$,$Z$) diagram.
In Fig.\ \ref{fig_dT_PM:CMDs} we overplotted the BT-Settl models
\citep{allard12} for the UKIDSS filter set \citep{hewett06} although our
survey uses the VISTA $ZYJ$ filters. However, those three VISTA filters 
are identical to the filters installed on UKIRT/WFCAM at the 1\% level
\footnote{http://apm49.ast.cam.ac.uk/surveys-projects/vista/technical/filter-set}. 

\subsection{Improving the photometric selection}
\label{USco_VISTA:select_optimal}

In this section we improve on our three-band photometric selection using
proper motions and near-infrared $HK$ photometry and proper motions from the 
UKIDSS GCS DR10 to confirm (or otherwise) photometric candidates identified 
in the previous section.

We selected point sources brighter than $J$\,=\,18.5 mag whose errors on 
the proper motions are within 4$\sigma$ of the mean USco proper motion 
(i.e.\ more than 99.99\% completeness), centered at ($-$6, $-$20) mas/yr. 
Fainter sources were kept because they will be fainter in UKIDSS than VISTA,
yielding large errors on the source centroid.
This selection yielded 66$+$17\,=\,83 candidates.

Second, we took advantage of the additional $H$- and $K$-band imaging from
the UKIDSS GCS DR10 to improve on our photometric and astrometric
selections. The ($Z-K$,$Z$) and ($J-K$,$J$) diagrams involving $Z$ and $J$
photometry from VISTA and $K$ photometry from GCS DR10 are displayed in 
Fig.\ \ref{fig_dT_PM:CMDs}. We applied the following photometric cuts
(dot-dashed lines in Fig.\ \ref{fig_dT_PM:CMDs}):
\begin{itemize}
\item ($Z-K$,$Z$) = (2.0,15.0) to (5.0,23.5)
\item ($J-K$,$J$) = (0.85,13.5) to (1.05,17.0)
\item ($J-K$,$J$) = (1.05,13.5) to (2.9,20.5)
\end{itemize}

\begin{figure}
  \centering
  \includegraphics[width=\linewidth, angle=0]{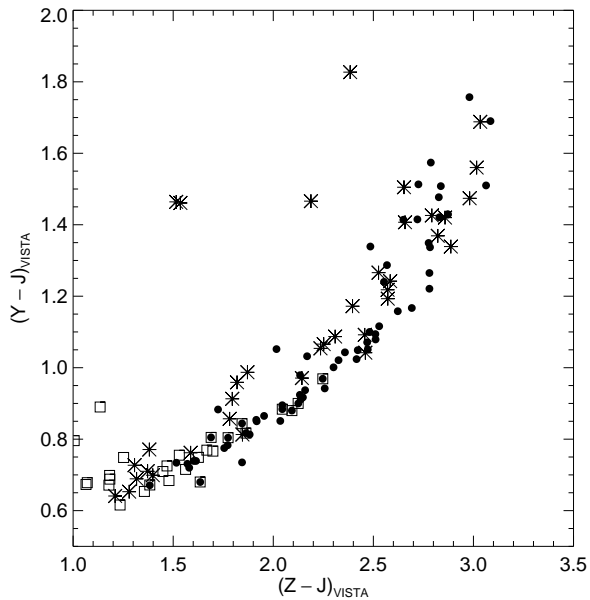}
  \caption{($Z-J$,$Y-J$) colour-colour diagram for USco member candidates
identified in the VISTA survey. Symbols as in Fig.\ \ref{fig_dT_PM:CMDs}}.
  \label{fig_dT_PM:CCD}
\end{figure}

This selection returned 67 candidates 
(Table \ref{tab_USco:final_candidates_VISTA}), which represent our final 
sample of bona-fide USco members in 13.5 square degrees down to a depth of 
$J$\,=\,20.5 mag, corresponding to $\sim$5 M$_{\rm Jup}$ according to
the BT-Settl models at 5 Myr \citep{allard12}.
We plot these new members as filled dots in the colour-magnitude and
colour-colour diagrams displayed in Figs \ref{fig_dT_PM:CMDs} and 
\ref{fig_dT_PM:CCD}. The candidates classified as 
photometric and/or astrometric non members are plotted as asterisks 
in Figs \ref{fig_dT_PM:CMDs}--\ref{fig_dT_PM:CCD} and listed in 
Table \ref{tab_USco:rejected_candidates_VISTA}.

We point out that the BT-Settl models at 5 Myr (or 10 Myr) have trouble
reproducing the observed USco sequence in the ($Z-J$,$Z$) colour-magnitude
diagram below the deuterium-burning limit. Moreover, the predicted $J-K$
colours are too blue, as seen in the ($J-K$,$J$) diagram
(Fig.\ \ref{fig_dT_PM:CMDs}). However, the kink predicted by the models 
at the deuterium-burning limit is clearly observed in the USco sequence,
demonstrating that we are clearly reaching sources with masses in the 
planetary-mass regime, independently of the inherent uncertainties
in the assignment of masses at young ages \citep{baraffe02,hillenbrand04}.

\subsection{$Z$-band drop-outs}
\label{USco_VISTA:select_Zdropouts}

We searched for $Z$-band drop-outs, selecting all sources 
lying to the right of a line running from ($Y-J$,$Y$)\,=\,(1.0, 20.0) 
to (1.75, 22.0) in that diagram. This query returned 30 candidates.
After inspecting the VISTA images, we rejected 25 of them because we can 
see a detection in the $Z$-band images at the position of the object in 
spite of the magnitude in this passband not being catalogued. The remaining
five candidates are listed in Table \ref{tab_USco:YJonly_candidates_VISTA}.

We found $K$-band magnitudes from the GCS DR10 for two of five $YJ$ 
candidates with $Z$ drop-outs (Table \ref{tab_USco:YJonly_candidates_VISTA}).
One of them (16:11:48.92$-$21:05:28.6) has two $K$-band measurements
(Table \ref{tab_USco:YJonly_candidates_VISTA}), indicating a possible 
intrinsic variability. Its $J-K$ colour span the 1.78--2.09 mag range, 
placing it close to the 10 Myr isochrone in the ($J-K$,$J$) diagram but 
to the blue of the cluster sequence. The other source with a $K$-band
detection would be bluer, with $J-K$\,=\,1.56 mag, suggesting that
it not a member unless the sequence turns to the blue if we reach the
L/T transition (models predict effective temperatures in the 1400--1200\,K 
range for a 5 M$_{\rm Jup}$ object at 5--10 Myr).
The other three candidates are undetected in the $K$-band, implying $J-K$ 
colours bluer than 1.6--1.8 mag assuming a 5$\sigma$ completeness limit of 
18.2 mag for the GCS $K$-band depth. 

We note that all five $YJ$ candidates lie outside the GTC/OSIRIS coverage 
discussed in \citet{lodieu13b}. However, we detected two of these five sources
in deep Magellan/IMACS data. One candidate (16:13:06.55$-$22:55:32.7, is clearly 
detected on the IMACS images at 0.145 arcsec from the VISTA detection.
We derived a magnitude of 21.878, implying a $Z-J$ colour of 2.08 mag
(Table \ref{tab_USco:YJonly_candidates_VISTA}),
placing this object clearly to the blue part of the cluster sequence.
Hence, we discard this object as a member of USco. The second candidate
(16:12:51.83$-$23:16:50.0) lies in the region of the detector which suffered 
strong vignetting, yielding large error bar on the photometry. We estimate 
its magnitude to 21.98--22.68, implying a $Z-J$ colour in the 2.14--2.79 mag
(Table \ref{tab_USco:YJonly_candidates_VISTA}).
This object also lies to the blue of the cluster sequence but we cannot
fully discard it until we obtain a better estimate of its magnitude. 
At this stage, we are unable to discard the other candidates because 
they could be L/T transition members in USco: deeper optical imaging and 
near-infrared spectroscopy can confirm this argument as well as their nature.

%
%
%

%
%
\begin{table*}
  \caption{Sample of 67 USco member candidates in the VISTA
deep survey: we list their coordinates, photometry, proper 
motions in both directions (in mas/yr) and their associated
error bars. The complete table is available in the electronic
version, only a subset is shown here for its content.
}
  \label{tab_USco:final_candidates_VISTA}
  \begin{tabular}{@{\hspace{0mm}}c @{\hspace{1.5mm}}c @{\hspace{1.5mm}}c @{\hspace{1.5mm}}c @{\hspace{1.5mm}}c @{\hspace{1.5mm}}c @{\hspace{1.5mm}}c @{\hspace{1.5mm}}c @{\hspace{1.5mm}}c @{\hspace{1.5mm}}c@{\hspace{0mm}}}
  \hline
R.A.\ & Dec.\  &  $Z$\,$\pm$\,err  &  $Y$\,$\pm$\,err  &  $J$\,$\pm$\,err  &  $H$\,$\pm$\,err  & $K$1\,$\pm$\,err & $K$2\,$\pm$\,err & $\mu_{\alpha}cos\delta$\,$\pm$\,err & $\mu_{\delta}$\,$\pm$\,err \cr
 \hline
15:59:36.38 & $-$22:14:15.9 & 19.523$\pm$0.014 & 18.210$\pm$0.013 & 16.970$\pm$0.009 & 16.341$\pm$0.028 & 15.672$\pm$0.024 & 15.638$\pm$0.015 &   $-$0.05$\pm$ 8.58 &  $-$19.40$\pm$ 8.58 \cr
15:59:48.01 & $-$22:27:16.5 & 15.899$\pm$0.002 & 14.919$\pm$0.002 & 14.144$\pm$0.001 & 13.610$\pm$0.003 & 13.125$\pm$0.003 & 13.147$\pm$0.002 &   $-$1.48$\pm$ 6.83 &  $-$14.09$\pm$ 6.83 \cr
 \ldots{}   & \ldots{}   & \ldots{} & \ldots{} & \ldots{} & \ldots{} & \ldots{} & \ldots{}  & \ldots{}  & \ldots{} \cr
16:17:00.53 & $-$22:12:53.9 & 18.170$\pm$0.007 & 16.778$\pm$0.004 & 15.754$\pm$0.004 & 15.084$\pm$0.008 & 14.419$\pm$0.007 & 14.468$\pm$0.005 &    2.37$\pm$ 6.99 &  $-$24.16$\pm$ 6.99 \cr
16:17:02.61 & $-$20:54:49.4 & 15.548$\pm$0.002 & 14.673$\pm$0.001 & 13.934$\pm$0.001 & 13.440$\pm$0.003 & 13.005$\pm$0.003 & 13.002$\pm$0.002 &   $-$3.39$\pm$ 4.44 &  $-$18.62$\pm$ 4.44 \cr
 \hline
\end{tabular}
\end{table*}
%

%
%
\begin{table*}
  \caption{Sample of 43 USco candidates classified as photometric
and/or astrometric non-members. This table provides their
coordinates, photometry, proper motions with their uncertainties.
The complete table is available in the electronic
version, only a subset is shown here for its content.
}
  \label{tab_USco:rejected_candidates_VISTA}
  \begin{tabular}{@{\hspace{0mm}}c @{\hspace{1.5mm}}c @{\hspace{1.5mm}}c @{\hspace{1.5mm}}c @{\hspace{1.5mm}}c @{\hspace{1.5mm}}c @{\hspace{1.5mm}}c @{\hspace{1.5mm}}c @{\hspace{1.5mm}}c @{\hspace{1.5mm}}c@{\hspace{0mm}}}
  \hline
R.A.\ & Dec.\  &  $Z$\,$\pm$\,err  &  $Y$\,$\pm$\,err  &  $J$\,$\pm$\,err  &  $H$\,$\pm$\,err  & $K$1\,$\pm$\,err & $K$2\,$\pm$\,err & $\mu_{\alpha}cos\delta$\,$\pm$\,err & $\mu_{\delta}$\,$\pm$\,err \cr
 \hline
15:59:37.56 & $-$22:54:16.3 & 16.338$\pm$0.003 & 15.166$\pm$0.002 & 14.195$\pm$0.001 & 13.554$\pm$0.003 & 13.007$\pm$0.002 & 13.011$\pm$0.002 & $-$225.61$\pm$ 6.82 & $-$214.84$\pm$ 6.82 \cr
15:59:57.99 & $-$21:54:22.4 & 22.410$\pm$0.159 & 20.903$\pm$0.109 & 19.429$\pm$0.074 & 19.427$\pm$0.331 & 18.963$\pm$0.385 & 17.848$\pm$0.126 &  $-$40.79$\pm$ 9.66 &  $-$29.17$\pm$ 9.66 \cr
 \ldots{}   & \ldots{}   & \ldots{} & \ldots{} & \ldots{} & \ldots{} & \ldots{} & \ldots{}  & \ldots{}  & \ldots{} \cr
16:18:25.46 & $-$23:40:53.5 & 17.797$\pm$0.007 & 16.915$\pm$0.005 & 16.002$\pm$0.004 & 15.051$\pm$0.007 & 14.553$\pm$0.008 & 14.525$\pm$0.006 &  $-$11.99$\pm$ 2.27 &   19.74$\pm$ 2.27 \cr
16:18:25.65 & $-$23:42:45.4 & 15.278$\pm$0.002 & 14.698$\pm$0.001 & 13.971$\pm$0.001 & 13.223$\pm$0.002 & 12.861$\pm$0.002 & 12.867$\pm$0.002 &   $-$3.73$\pm$ 2.19 &   $-$1.63$\pm$ 2.19 \cr
 \hline
\end{tabular}
\end{table*}
%

%
%
\begin{table}
  \centering
  \caption{$Z$-band drop-outs (or $YJ$-only) candidates in USco. The $K$-band
photometry is from the UKIDSS GCS DR10\@. $^{a}$ We list the $K$-band magnitude
with the lowest error bar: the other measurement available from GCS DR10 gives
$K$\,=\,17.579$\pm$0.133 mag. $^{b}$ This object, covered by
our IMACS survey, has a magnitude of 21.878 ($Z-J$\,=\,20.8 mag) and is rejected
as a member. $^{c}$ This candidate lies in the vignetted part of the IMACS detector.
We estimated its $Z-J$ colour to 2.14--2.79 mag.
}
  \label{tab_USco:YJonly_candidates_VISTA}
  \begin{tabular}{@{\hspace{0mm}}c @{\hspace{2mm}}c @{\hspace{2mm}}c @{\hspace{2mm}}c @{\hspace{2mm}}c@{\hspace{0mm}}}
  \hline
R.A.\       & Dec.\               &  $Y$\,$\pm$\,err  &  $J$\,$\pm$\,err & $K$\,$\pm$\,err \cr
 \hline
hh:mm:ss.ss & ${^\circ}$:$'$:$''$ &    mag            &     mag    & mag   \cr 
 \hline
16:11:48.92 & $-$21:05:28.6 & 21.173$\pm$0.152 & 19.669$\pm$0.096 & 17.887$\pm$0.108$^{a}$ \cr
16:13:06.55 & $-$22:55:32.7 & 21.303$\pm$0.155 & 19.798$\pm$0.090 & $>$18.2$^{b}$ \cr
16:12:51.83 & $-$23:16:50.0 & 21.437$\pm$0.203 & 19.838$\pm$0.095 & $>$18.2$^{c}$ \cr
16:16:21.04 & $-$23:55:20.1 & 21.624$\pm$0.241 & 19.998$\pm$0.109 & 18.438$\pm$0.167 \cr
16:10:27.63 & $-$23:05:54.3 & 21.767$\pm$0.438 & 20.004$\pm$0.145 & $>$18.2 \cr
 \hline
\end{tabular}
\end{table}
%

%
%
\section{Discussion on the mass function}
\label{USco_VISTA:MF}

Before looking at the shape of the USco mass function in the planetary-mass
regime, we discuss the mass estimates and place our work into context.

Masses at young ages ($<$10 Myr) are difficult to measure and can be very 
uncertain \citep{baraffe02} and depend also on the models considered during
the analysis \citep{hillenbrand04}. The lack of spectroscopy for the lowest 
and youngest brown dwarf candidates identified photometrically to date in 
star-forming regions and clusters hampers our ability to estimate masses accurately.
To estimate the depth of our VISTA survey, we rely on the latest BT-Settl
models \citep{allard12}, assuming a distance of 145 pc and ages of 5 and
10 Myr for USco. According to these models, a 5 M$_{\rm Jup}$ brown dwarf
with an effective temperature of 1400--1200\,K in USco has apparent 
magnitudes of $Z$\,$\sim$\,23.1--23.9 mag, 
$Y$\,$\sim$\,21.4--22.1 mag, and $J$\,$\sim$\,19.8--20.6 mag, respectively
(depending on the age chosen). These $J$-band values are consistent with our
100\% completeness limit so we are sensitive to 5 M$_{\rm Jup}$ brown dwarfs 
in USco. However, the $Y$- and $Z$-band magnitude of a 5 M$_{\rm Jup}$ brown 
dwarf in USco are fainter than our completeness limits, implying that we
are complete down to 6 M$_{\rm Jup}$ ($Y$\,=\,20.16 mag and $Z$\,=\,21.79 mag).

In terms of magnitude, our survey is shallower than the deep $ZYJHK_{s}$ VISTA 
survey in $\sigma$ Orionis ($J$\,=\,21 mag vs 20.5 mag) by \citet{penya12a}.
However, both surveys have very similar depths in terms of mass because 
$\sigma$\,Ori is further away than USco (352 pc vs 145 pc), resulting in a 
5 M$_{\rm Jup}$ having $J$\,=\,20.53 mag in $\sigma$\,Ori and $J$\,=\,20.57 Myr 
for USco, assuming 10 Myr ($J$\,=\,19.77 mag at 5 Myr in USco). The coolest 
spectroscopically-confirmed member of $\sigma$\,Ori, sOri\,65
\citep[$Y$\,=\,21.1$\pm$0.1; $J$\,=\,20.3$\pm$0.1 mag; L3.5;][]{zapatero00,sherry04,scholz08a,penya12a}
is close to our 100\% completeness so we predict that our VISTA survey
is sensitive to mid-L dwarfs in USco, considering the similarities between
both surveys. Nonetheless, we cannot discard that we are sensitive to L/T
transition objects because of the similar apparent 
magnitudes observed for field brown dwarfs across the mid-L to mid-T regime 
\citep[e.g.][]{leggett10a}.

To address the shape of the USco mass function in the planetary-mass regime
\citep{salpeter55,miller79,scalo86}, we divided our sample into two mass 
bins of equal size in a logarithmic scale. We counted the number of candidates
in the 20--10 M$_{\rm Jup}$ and 10--5 M$_{\rm Jup}$ bins, resulting in a ratio 
of 24/22\,=\,1.09 (27/16\,=\,1.69), respectively, using the $J$-band filter 
and assuming an age of 5 Myr for USco (values in brackets are for 10 Myr).
If we consider the $Y$-band filter instead, the ratios become 25/21\,=\,1.19
(5 Myr) and 30/16\,=\,1.88 (10 Myr), respectively. These ratios are upper limits 
because our survey is complete down to 6 M$_{\rm Jup}$ and incomplete in the 
6--5 M$_{\rm Jup}$ interval. Therefore, we conclude that the USco mass
function is more likely to decrease in the planetary-mass regime although
a flat mass function cannot be discarded for an age of 5 Myr until proper 
motions or near-infrared
spectroscopy is obtained for all candidates. Our conclusions agree with the
possible turn-over of the mass function in the planetary-mass regime observed
in $\sigma$\,Ori \citep{penya12a} and NGC\,1333 \citep{scholz12c} but at
odds with the rising mass function seen in $\rho$\,Oph \citep{marsh10,barsony12}.

Our future goal would
be to reach 1 M$_{\rm Jup}$ to test the various theoretical models proposing
discrepant minimum masses: self-gravitating disk fragmentation or
gravitationally unstable disks lead to minimum masses of 3--5 M$_{\rm Jup}$ 
whereas fragmentation of a shock-compressed layer suggests minimum masses 
below 3 M$_{\rm Jup}$, in agreement with the original prediction by 
\citet{kumar69}. The latest evolutionary BT-Settl models predict absolute
$J$ magnitudes 1.5 to 2 mag fainter for a 3 M$_{\rm Jup}$ compared to a 
5 M$_{\rm Jup}$ brown dwarfs at 5 or 10 Myr, consistent with differences
predicted by the COND models \citep{baraffe02}. Similarly, the COND models
predict that a 1 M$_{\rm Jup}$ object in USco would be 3 mag fainter
than a 3 M$_{\rm Jup}$ in $J$, implying that we would need a survey down to
$J$\,$\sim$\,25 mag, depth very difficult to achieve from the ground
over a wide area (at least six square degrees are needed for statistical 
purposes)\footnote{The 5-year UltraVISTA public survey aims at reaching
5$\sigma$ depths of $Y$\,$\sim$\,26 mag, $J$\,$\sim$25.5 mag in the Vega
system over 1.5 square degrees \citep{mccracken12}}. Although not ideal 
for large-scale surveys, near and mid-infrared 
capabilities of the James Webb Space Telescope seem the only option to 
address the issue of the fragmentation limit by the end of the decade.

%
%
\section*{Acknowledgments}
NL was funded by the Ram\'on y Cajal fellowship number 08-303-01-02
and the national program AYA2010-19136 funded by the Spanish ministry 
of Science and Innovation. 

This work is based on observations made with the ESO VISTA telescope
at the Paranal Observatory under programme ID 089.C-0102(ABC) in service 
mode and in visitor mode with the 6.5-m Magellan telescope at
Las Campanas Observatory Carnegie Institution of Washington.

This work is based in part on data obtained as part of the UKIRT
Infrared Deep Sky Survey (UKIDSS). 
We thank our colleagues at the UK Astronomy Technology Centre, the Joint
Astronomy Centre in Hawaii, the Cambridge Astronomical Survey and Edinburgh
Wide Field Astronomy Units for building and operating WFCAM and its
associated data flow system.

This research has made use of the Simbad and Vizier \citep{ochsenbein00}
databases, operated at the Centre de Donn\'ees Astronomiques de Strasbourg 
(CDS), and of NASA's Astrophysics Data System Bibliographic Services (ADS). 

%
%
\bibliographystyle{mn2e}
\bibliography{../../AA/mnemonic,../../AA/biblio_old}

\begin{thebibliography}{}

\bibitem[\protect\citeauthoryear{{Allard}, {Homeier} \& {Freytag}}{{Allard}
  et~al.}{2012}]{allard12}
{Allard} F.,  {Homeier} D.,    {Freytag} B.,  2012, Royal Society of London
  Philosophical Transactions Series A, 370, 2765

\bibitem[\protect\citeauthoryear{{Alves de Oliveira}, {Moraux}, {Bouvier} \&
  {Bouy}}{{Alves de Oliveira} et~al.}{2012}]{alves12}
{Alves de Oliveira} C.,  {Moraux} E.,  {Bouvier} J.,    {Bouy} H.,  2012, A\&A,
  539, A151

\bibitem[\protect\citeauthoryear{{Alves de Oliveira}, {Moraux}, {Bouvier},
  {Duch{\^e}ne}, {Bouy}, {Maschberger} \& {Hudelot}}{{Alves de Oliveira}
  et~al.}{2013}]{alves13a}
{Alves de Oliveira} C.,  {Moraux} E.,  {Bouvier} J.,  {Duch{\^e}ne} G.,  {Bouy}
  H.,  {Maschberger} T.,    {Hudelot} P.,  2013, A\&A, 549, A123

\bibitem[\protect\citeauthoryear{{Ardila}, {Mart{\'{\i}}n} \& {Basri}}{{Ardila}
  et~al.}{2000}]{ardila00}
{Ardila} D.,  {Mart{\'{\i}}n} E.,    {Basri} G.,  2000, AJ, 120, 479

\bibitem[\protect\citeauthoryear{{Baraffe}, {Chabrier}, {Allard} \&
  {Hauschildt}}{{Baraffe} et~al.}{1998}]{baraffe98}
{Baraffe} I.,  {Chabrier} G.,  {Allard} F.,    {Hauschildt} P.~H.,  1998, A\&A,
  337, 403

\bibitem[\protect\citeauthoryear{{Baraffe}, {Chabrier}, {Allard} \&
  {Hauschildt}}{{Baraffe} et~al.}{2002}]{baraffe02}
{Baraffe} I.,  {Chabrier} G.,  {Allard} F.,    {Hauschildt} P.~H.,  2002, A\&A,
  382, 563

\bibitem[\protect\citeauthoryear{{Barrado y Navascu{\'e}s}, {Zapatero Osorio},
  {B{\' e}jar}, {Rebolo}, {Mart{\'{\i}}n}, {Mundt} \& {Bailer-Jones}}{{Barrado
  y Navascu{\'e}s} et~al.}{2001}]{barrado01c}
{Barrado y Navascu{\'e}s} D.,  {Zapatero Osorio} M.~R.,  {B{\' e}jar} V.~J.~S.,
   {Rebolo} R.,  {Mart{\'{\i}}n} E.~L.,  {Mundt} R.,    {Bailer-Jones}
  C.~A.~L.,  2001, A\&A, 377, L9

\bibitem[\protect\citeauthoryear{{Barsony}, {Haisch}, {Marsh} \&
  {McCarthy}}{{Barsony} et~al.}{2012}]{barsony12}
{Barsony} M.,  {Haisch} K.~E.,  {Marsh} K.~A.,    {McCarthy} C.,  2012, ApJ,
  751, 22

\bibitem[\protect\citeauthoryear{{Basri}, {Mohanty}, {Allard}, {Hauschildt},
  {Delfosse}, {Mart{\'{\i}}n}, {Forveille} \& {Goldman}}{{Basri}
  et~al.}{2000}]{basri00}
{Basri} G.,  {Mohanty} S.,  {Allard} F.,  {Hauschildt} P.~H.,  {Delfosse} X.,
  {Mart{\'{\i}}n} E.~L.,  {Forveille} T.,    {Goldman} B.,  2000, ApJ, 538, 363

\bibitem[\protect\citeauthoryear{{Bastian}, {Covey} \& {Meyer}}{{Bastian}
  et~al.}{2010}]{bastian10}
{Bastian} N.,  {Covey} K.~R.,    {Meyer} M.~R.,  2010, ARA\&A, 48, 339

\bibitem[\protect\citeauthoryear{{Bihain}, {Rebolo}, {Zapatero Osorio},
  {B{\'e}jar}, {Vill{\'o}-P{\'e}rez}, {D{\'{\i}}az-S{\'a}nchez},
  {P{\'e}rez-Garrido} \& {9 co-authors}}{{Bihain} et~al.}{2009}]{bihain09}
{Bihain} G., et al.\ 2009, A\&A, 506, 1169

\bibitem[\protect\citeauthoryear{{Boss}}{{Boss}}{1988}]{boss88}
{Boss} A.~P.,  1988, ApJ, 331, 370

\bibitem[\protect\citeauthoryear{{Boudreault} \& {Lodieu}}{{Boudreault} \&
  {Lodieu}}{2013}]{boudreault13}
{Boudreault} S.,  {Lodieu} N.,  2013, MNRAS

\bibitem[\protect\citeauthoryear{{Bouvier}, {Kendall}, {Meeus}, {Testi},
  {Moraux}, {Stauffer}, {James}, {Cuillandre}, {Irwin}, {McCaughrean},
  {Baraffe} \& {Bertin}}{{Bouvier} et~al.}{2008}]{bouvier08a}
{Bouvier} J., et al.\ 2008, A\&A, 481, 661

\bibitem[\protect\citeauthoryear{{Boyd} \& {Whitworth}}{{Boyd} \&
  {Whitworth}}{2005}]{boyd05}
{Boyd} D.~F.~A.,  {Whitworth} A.~P.,  2005, A\&A, 430, 1059

\bibitem[\protect\citeauthoryear{{Burgasser}, {Geballe}, {Leggett},
  {Kirkpatrick} \& {Golimowski}}{{Burgasser} et~al.}{2006}]{burgasser06a}
{Burgasser} A.~J.,  {Geballe} T.~R.,  {Leggett} S.~K.,  {Kirkpatrick} J.~D.,
  {Golimowski} D.~A.,  2006, ApJ, 637, 1067

\bibitem[\protect\citeauthoryear{{Burgess}, {Moraux}, {Bouvier}, {Marmo},
  {Albert} \& {Bouy}}{{Burgess} et~al.}{2009}]{burgess09}
{Burgess} A.~S.~M.,  {Moraux} E.,  {Bouvier} J.,  {Marmo} C.,  {Albert} L.,
  {Bouy} H.,  2009, A\&A, 508, 823

\bibitem[\protect\citeauthoryear{{Caballero}, {B{\'e}jar}, {Rebolo},
  {Eisl{\"o}ffel}, {Zapatero Osorio}, {Mundt}, {Barrado Y Navascu{\'e}s},
  {Bihain}, {Bailer-Jones}, {Forveille} \& {Mart{\'{\i}}n}}{{Caballero}
  et~al.}{2007}]{caballero07d}
{Caballero} J.~A.,  et al.\ 2007, A\&A, 470, 903

\bibitem[\protect\citeauthoryear{{Casali}, {Adamson}, {Alves de Oliveira},
  {Almaini}, {Burch}, {Chuter}, {Elliot} \& {23 co-authors}}{{Casali}
  et~al.}{2007}]{casali07}
{Casali} M., et al.\ 2007, A\&A, 467, 777

\bibitem[\protect\citeauthoryear{{Casewell}, {Dobbie}, {Hodgkin}, {Moraux},
  {Jameson}, {Hambly}, {Irwin} \& {Lodieu}}{{Casewell}
  et~al.}{2007}]{casewell07}
{Casewell} S.~L.,  {Dobbie} P.~D.,  {Hodgkin} S.~T.,  {Moraux} E.,  {Jameson}
  R.~F.,  {Hambly} N.~C.,  {Irwin} J.,    {Lodieu} N.,  2007, MNRAS, 378, 1131

\bibitem[\protect\citeauthoryear{{Chabrier}}{{Chabrier}}{2003}]{chabrier03}
{Chabrier} G.,  2003, PASP, 115, 763

\bibitem[\protect\citeauthoryear{{Chabrier}}{{Chabrier}}{2005}]{chabrier05a}
{Chabrier} G.,  2005, in {E.~Corbelli, F.~Palla, \& H.~Zinnecker} ed., The
  Initial Mass Function 50 Years Later Vol.~327 of Astrophysics and Space
  Science Library, {The Initial Mass Function: from Salpeter 1955 to 2005}.
p.~41

\bibitem[\protect\citeauthoryear{{Chabrier}, {Baraffe}, {Allard} \&
  {Hauschildt}}{{Chabrier} et~al.}{2000}]{chabrier00c}
{Chabrier} G.,  {Baraffe} I.,  {Allard} F.,    {Hauschildt} P.,  2000, ApJ,
  542, 464

\bibitem[\protect\citeauthoryear{{Collins} \& {Hambly}}{{Collins} \&
  {Hambly}}{2012}]{collins12}
{Collins} R.,  {Hambly} N.,  2012, in "Ballester P.,  Egret D.,  eds,
  Astronomical Data Analysis Software and Systems XXI Vol. in press, of
  Astronomical Society of the Pacific Conference Series, {Calculating proper
  motions in the WFCAM Science Archive for the UKIRT Infrared Deep Sky Surveys}

\bibitem[\protect\citeauthoryear{{Cross}, {Collins}, {Mann}, {Read},
  {Sutorius}, {Blake}, {Holliman}, {Hambly}, {Emerson}, {Lawrence} \&
  {Noddle}}{{Cross} et~al.}{2012}]{cross12}
{Cross} N.~J.~G., et al.\ 2012, A\&A, 548, A119

\bibitem[\protect\citeauthoryear{{Cutri}, {Skrutskie}, {van Dyk}, {Beichman},
  {Carpenter}, {Chester}, {Cambresy}, {Evans}, {Fowler}, {Gizis} \& {15
  coauthors}}{{Cutri} et~al.}{2003}]{cutri03}
{Cutri} R.~M., et al.\ 2003, 2MASS All Sky Catalog of point sources, 2246

\bibitem[\protect\citeauthoryear{{Dalton}, {Caldwell}, {Ward}, {Whalley},
  {Woodhouse}, {Edeson}, {Clark}, {Beard}, {Gallie}, {Todd}, {Strachan},
  {Bezawada}, {Sutherland} \& {Emerson}}{{Dalton} et~al.}{2006}]{dalton06}
{Dalton} G.~B., et al.\ 2006, in Society of Photo-Optical Instrumentation Engineers
  (SPIE) Conference Series Vol.~6269 of Presented at the Society of
  Photo-Optical Instrumentation Engineers (SPIE) Conference, {The VISTA
  infrared camera}

\bibitem[\protect\citeauthoryear{{Dawson}, {Scholz} \& {Ray}}{{Dawson}
  et~al.}{2011}]{dawson11}
{Dawson} P.,  {Scholz} A.,    {Ray} T.~P.,  2011, A\&A

\bibitem[\protect\citeauthoryear{{Dawson}, {Scholz}, {Ray}, {Marsh}, {Wood},
  {Natta}, {Padgett} \& {Ressler}}{{Dawson} et~al.}{2012}]{dawson12}
{Dawson} P.,  {Scholz} A.,  {Ray} T.~P.,  {Marsh} K.~A.,  {Wood} K.,  {Natta}
  A.,  {Padgett} D.,    {Ressler} M.~E.,  2012, MNRAS

\bibitem[\protect\citeauthoryear{{de Bruijne}, {Hoogerwerf}, {Brown}, {Aguilar}
  \& {de Zeeuw}}{{de Bruijne} et~al.}{1997}]{deBruijne97}
{de Bruijne} J.~H.~J.,  {Hoogerwerf} R.,  {Brown} A.~G.~A.,  {Aguilar} L.~A.,
   {de Zeeuw} P.~T.,  1997, in ESA SP-402: Hipparcos - Venice '97 {Improved
  Methods for Identifying Moving Groups}.
pp 575--578

\bibitem[\protect\citeauthoryear{{de Zeeuw}, {Hoogerwerf}, {de Bruijne},
  {Brown} \& {Blaauw}}{{de Zeeuw} et~al.}{1999}]{deZeeuw99}
{de Zeeuw} P.~T.,  {Hoogerwerf} R.,  {de Bruijne} J.~H.~J.,  {Brown} A.~G.~A.,
    {Blaauw} A.,  1999, AJ, 117, 354

\bibitem[\protect\citeauthoryear{{Dye}, {Warren}, {Hambly}, {Cross}, {Hodgkin},
  {Irwin}, {Lawrence}, {Adamson} \& {37 co-authors}}{{Dye}
  et~al.}{2006}]{dye06}
{Dye} S., et al.\ 2006, MNRAS, 372, 1227

\bibitem[\protect\citeauthoryear{{Emerson}}{{Emerson}}{2001}]{emerson01}
{Emerson} J.~P.,  2001, in {R.~Clowes, A.~Adamson, \& G.~Bromage} ed., The New
  Era of Wide Field Astronomy Vol.~232 of Astronomical Society of the Pacific
  Conference Series, {VISTA - Project Status of the Visible and Infrared Survey
  Telescope for Astronomy.}.
p.~339

\bibitem[\protect\citeauthoryear{{Emerson}, {Sutherland}, {McPherson}, {Craig},
  {Dalton} \& {Ward}}{{Emerson} et~al.}{2004}]{emerson04}
{Emerson} J.~P.,  {Sutherland} W.~J.,  {McPherson} A.~M.,  {Craig} S.~C.,
  {Dalton} G.~B.,    {Ward} A.~K.,  2004, The Messenger, 117, 27

\bibitem[\protect\citeauthoryear{{Forgan} \& {Rice}}{{Forgan} \&
  {Rice}}{2011}]{forgan11}
{Forgan} D.,  {Rice} K.,  2011, MNRAS, 417, 1928

\bibitem[\protect\citeauthoryear{{Fossati}, {Bagnulo}, {Landstreet}, {Wade},
  {Kochukhov}, {Monier}, {Weiss} \& {Gebran}}{{Fossati}
  et~al.}{2008}]{fossati08}
{Fossati} L.,  {Bagnulo} S.,  {Landstreet} J.,  {Wade} G.,  {Kochukhov} O.,
  {Monier} R.,  {Weiss} W.,    {Gebran} M.,  2008, A\&A, 483, 891

\bibitem[\protect\citeauthoryear{{Hambly}, {Collins}, {Cross}, {Mann}, {Read},
  {Sutorius}, {Bond}, {Bryant}, {Emerson}, {Lawrence}, {Rimoldini}, {Stewart},
  {Williams}, {Adamson}, {Hirst}, {Dye} \& {Warren}}{{Hambly}
  et~al.}{2008}]{hambly08}
{Hambly} N.~C., et al.\ 2008, MNRAS, 384, 637

\bibitem[\protect\citeauthoryear{{Hambly}, {MacGillivray}, {Read}, {Tritton},
  {Thomson}, {Kelly}, {Morgan}, {Smith}, {Driver}, {Williamson}, {Parker},
  {Hawkins}, {Williams} \& {Lawrence}}{{Hambly} et~al.}{2001}]{hambly01a}
{Hambly} N.~C., et al.\ 2001, MNRAS, 326, 1279

\bibitem[\protect\citeauthoryear{{Hewett}, {Warren}, {Leggett} \&
  {Hodgkin}}{{Hewett} et~al.}{2006}]{hewett06}
{Hewett} P.~C.,  {Warren} S.~J.,  {Leggett} S.~K.,    {Hodgkin} S.~T.,  2006,
  MNRAS, 367, 454

\bibitem[\protect\citeauthoryear{{Hillenbrand} \& {White}}{{Hillenbrand} \&
  {White}}{2004}]{hillenbrand04}
{Hillenbrand} L.~A.,  {White} R.~J.,  2004, ApJ, 604, 741

\bibitem[\protect\citeauthoryear{{Hodgkin}, {Irwin}, {Hewett} \&
  {Warren}}{{Hodgkin} et~al.}{2009}]{hodgkin09}
{Hodgkin} S.~T.,  {Irwin} M.~J.,  {Hewett} P.~C.,    {Warren} S.~J.,  2009,
  MNRAS, 394, 675

\bibitem[\protect\citeauthoryear{{Hogan}, {Jameson}, {Casewell}, {Osbourne} \&
  {Hambly}}{{Hogan} et~al.}{2008}]{hogan08}
{Hogan} E.,  {Jameson} R.~F.,  {Casewell} S.~L.,  {Osbourne} S.~L.,    {Hambly}
  N.~C.,  2008, MNRAS, 388, 495

\bibitem[\protect\citeauthoryear{{Irwin} \& {Lewis}}{{Irwin} \&
  {Lewis}}{2001}]{irwin01}
{Irwin} M.,  {Lewis} J.,  2001, New Astronomy Review, 45, 105

\bibitem[\protect\citeauthoryear{{Irwin}, {Lewis}, {Hodgkin}, {Bunclark},
  {Evans}, {McMahon}, {Emerson}, {Stewart} \& {Beard}}{{Irwin}
  et~al.}{2004}]{irwin04}
{Irwin} M.~J., et al.\ 2004, in
  {Quinn} P.~J.,  {Bridger} A.,  eds, Optimizing Scientific Return for
  Astronomy through Information Technologies. Edited by Quinn, Peter J.;
  Bridger, Alan. Proceedings of the SPIE, Volume 5493, pp. 411-422 (2004).
  {VISTA data flow system: pipeline processing for WFCAM and VISTA}.
pp 411--422

\bibitem[\protect\citeauthoryear{{Kirkpatrick}}{{Kirkpatrick}}{2005}]{kirkpatrick05}
{Kirkpatrick} J.~D.,  2005, ARA\&A, 43, 195

\bibitem[\protect\citeauthoryear{{Kroupa}}{{Kroupa}}{2002}]{kroupa02}
{Kroupa} P.,  2002, Science, 295, 82

\bibitem[\protect\citeauthoryear{{Kumar}}{{Kumar}}{1969}]{kumar69}
{Kumar} S.~S.,  1969, in {Kumar} S.~S.,  ed., Low-Luminosity Stars {The nature
  of low-mass 'dark' companions}.
p.~255

\bibitem[\protect\citeauthoryear{{Kunkel}}{{Kunkel}}{1999}]{kunkel99}
{Kunkel} M.,  1999, Ph.D.~Thesis, Julius-Maximilians-Universit\"at W\"urzburg

\bibitem[\protect\citeauthoryear{{Lawrence}, {Warren}, {Almaini}, {Edge},
  {Hambly} \& {17 co-authors}}{{Lawrence} et~al.}{2007}]{lawrence07}
{Lawrence} A., et al.\ 2007, MNRAS, 379, 1599

\bibitem[\protect\citeauthoryear{{Leggett}, {Burningham}, {Saumon}, {Marley},
  {Warren}, {Smart}, {Jones}, {Lucas}, {Pinfield} \& {Tamura}}{{Leggett}
  et~al.}{2010}]{leggett10a}
{Leggett} S.~K., et al.\ 2010, ApJ, 710, 1627

\bibitem[\protect\citeauthoryear{{Leggett}, {Geballe}, {Fan}, {Schneider},
  {Gunn}, {Lupton}, {Knapp}, {Strauss} \& {27 coauthors}}{{Leggett}
  et~al.}{2000}]{leggett00}
{Leggett} S.~K., et al.\ 2000, ApJL, 536, L35

\bibitem[\protect\citeauthoryear{{Lodieu}}{{Lodieu}}{2013}]{lodieu13c}
{Lodieu} N.,  2013, MNRAS

\bibitem[\protect\citeauthoryear{{Lodieu}, {Deacon} \& {Hambly}}{{Lodieu}
  et~al.}{2012}]{lodieu12a}
{Lodieu} N.,  {Deacon} N.~R.,    {Hambly} N.~C.,  2012, MNRAS, p.~2699

\bibitem[\protect\citeauthoryear{{Lodieu}, {Dobbie} \& {Hambly}}{{Lodieu}
  et~al.}{2011}]{lodieu11a}
{Lodieu} N.,  {Dobbie} P.~D.,    {Hambly} N.~C.,  2011, A\&A, 527, A24

\bibitem[\protect\citeauthoryear{{Lodieu}, {Hambly}, {Dobbie}, {Cross},
  {Christensen}, {Martin} \& {Valdivielso}}{{Lodieu} et~al.}{2011}]{lodieu11c}
{Lodieu} N.,  {Hambly} N.~C.,  {Dobbie} P.~D.,  {Cross} N.~J.~G.,
  {Christensen} L.,  {Martin} E.~L.,    {Valdivielso} L.,  2011, MNRAS, 418,
  2604

\bibitem[\protect\citeauthoryear{{Lodieu}, {Hambly} \& {Jameson}}{{Lodieu}
  et~al.}{2006}]{lodieu06}
{Lodieu} N.,  {Hambly} N.~C.,    {Jameson} R.~F.,  2006, MNRAS, 373, 95

\bibitem[\protect\citeauthoryear{{Lodieu}, {Hambly}, {Jameson}, {Hodgkin},
  {Carraro} \& {Kendall}}{{Lodieu} et~al.}{2007}]{lodieu07a}
{Lodieu} N.,  {Hambly} N.~C.,  {Jameson} R.~F.,  {Hodgkin} S.~T.,  {Carraro}
  G.,    {Kendall} T.~R.,  2007, MNRAS, 374, 372

\bibitem[\protect\citeauthoryear{{Lodieu}, {Ivanov} \& {Dobbie}}{{Lodieu}
  et~al.}{2013}]{lodieu13b}
{Lodieu} N.,  {Ivanov} V.~D.,    {Dobbie} P.~D.,  2013, MNRAS, 430, 1784

\bibitem[\protect\citeauthoryear{{Low} \& {Lynden-Bell}}{{Low} \&
  {Lynden-Bell}}{1976}]{low76}
{Low} C.,  {Lynden-Bell} D.,  1976, MNRAS, 176, 367

\bibitem[\protect\citeauthoryear{{Marsh}, {Plavchan}, {Kirkpatrick},
  {Lowrance}, {Cutri} \& {Velusamy}}{{Marsh} et~al.}{2010}]{marsh10}
{Marsh} K.~A.,  {Plavchan} P.,  {Kirkpatrick} J.~D.,  {Lowrance} P.~J.,
  {Cutri} R.~M.,    {Velusamy} T.,  2010, ApJ, 719, 550

\bibitem[\protect\citeauthoryear{{Mart{\'{\i}}n}, {Delfosse} \&
  {Guieu}}{{Mart{\'{\i}}n} et~al.}{2004}]{martin04}
{Mart{\'{\i}}n} E.~L.,  {Delfosse} X.,    {Guieu} S.,  2004, AJ, 127, 449

\bibitem[\protect\citeauthoryear{{Mart{\'{\i}}n}, {Zapatero Osorio}, {Barrado y
  Navascu{\' e}s}, {B{\' e}jar} \& {Rebolo}}{{Mart{\'{\i}}n}
  et~al.}{2001}]{martin01a}
{Mart{\'{\i}}n} E.~L.,  {Zapatero Osorio} M.~R.,  {Barrado y Navascu{\' e}s}
  D.,  {B{\' e}jar} V.~J.~S.,    {Rebolo} R.,  2001, ApJL, 558, L117

\bibitem[\protect\citeauthoryear{{McCracken}, {Milvang-Jensen}, {Dunlop},
  {Franx}, {Fynbo}, {Le F{\`e}vre}, {Holt}, {Caputi} \& {13
  co-authors}}{{McCracken} et~al.}{2012}]{mccracken12}
{McCracken} H.~J., et al.\ 2012, A\&A, 544, A156

\bibitem[\protect\citeauthoryear{{Miller} \& {Scalo}}{{Miller} \&
  {Scalo}}{1979}]{miller79}
{Miller} G.~E.,  {Scalo} J.~M.,  1979, ApJS, 41, 513

\bibitem[\protect\citeauthoryear{{Ochsenbein}, {Bauer} \&
  {Marcout}}{{Ochsenbein} et~al.}{2000}]{ochsenbein00}
{Ochsenbein} F.,  {Bauer} P.,    {Marcout} J.,  2000, A\&AS, 143, 23

\bibitem[\protect\citeauthoryear{{Pe{\~n}a Ram{\'{\i}}rez}, {B{\'e}jar},
  {Zapatero Osorio}, {Petr-Gotzens} \& {Mart{\'{\i}}n}}{{Pe{\~n}a
  Ram{\'{\i}}rez} et~al.}{2012}]{penya12a}
{Pe{\~n}a Ram{\'{\i}}rez} K.,  {B{\'e}jar} V.~J.~S.,  {Zapatero Osorio} M.~R.,
  {Petr-Gotzens} M.~G.,    {Mart{\'{\i}}n} E.~L.,  2012, ApJ, 754, 30

\bibitem[\protect\citeauthoryear{{Pe{\~n}a Ram{\'{\i}}rez}, {Zapatero Osorio},
  {B{\'e}jar}, {Rebolo} \& {Bihain}}{{Pe{\~n}a Ram{\'{\i}}rez}
  et~al.}{2011}]{penya11a}
{Pe{\~n}a Ram{\'{\i}}rez} K.,  {Zapatero Osorio} M.~R.,  {B{\'e}jar} V.~J.~S.,
  {Rebolo} R.,    {Bihain} G.,  2011, A\&A, 532, A42

\bibitem[\protect\citeauthoryear{{Pecaut}, {Mamajek} \& {Bubar}}{{Pecaut}
  et~al.}{2012}]{pecaut12}
{Pecaut} M.~J.,  {Mamajek} E.~E.,    {Bubar} E.~J.,  2012, ApJ, 746, 154

\bibitem[\protect\citeauthoryear{{Perryman}, {Brown}, {Lebreton}, {Gomez},
  {Turon}, {de Strobel}, {Mermilliod}, {Robichon}, {Kovalevsky} \&
  {Crifo}}{{Perryman} et~al.}{1998}]{perryman98}
{Perryman} M.~A.~C.,  et al.\ 1998, A\&A, 331, 81

\bibitem[\protect\citeauthoryear{{Preibisch}, {Guenther}, {Zinnecker},
  {Sterzik}, {Frink} \& {Roeser}}{{Preibisch} et~al.}{1998}]{preibisch98}
{Preibisch} T.,  {Guenther} E.,  {Zinnecker} H.,  {Sterzik} M.,  {Frink} S.,
  {Roeser} S.,  1998, A\&A, 333, 619

\bibitem[\protect\citeauthoryear{{Preibisch} \& {Zinnecker}}{{Preibisch} \&
  {Zinnecker}}{1999}]{preibisch99}
{Preibisch} T.,  {Zinnecker} H.,  1999, AJ, 117, 2381

\bibitem[\protect\citeauthoryear{{Preibisch} \& {Zinnecker}}{{Preibisch} \&
  {Zinnecker}}{2002}]{preibisch02}
{Preibisch} T.,  {Zinnecker} H.,  2002, AJ, 123, 1613

\bibitem[\protect\citeauthoryear{{Rees}}{{Rees}}{1976}]{rees76}
{Rees} M.~J.,  1976, MNRAS, 176, 483

\bibitem[\protect\citeauthoryear{{Rogers} \& {Wadsley}}{{Rogers} \&
  {Wadsley}}{2012}]{rogers12}
{Rogers} P.~D.,  {Wadsley} J.,  2012, MNRAS, 423, 1896

\bibitem[\protect\citeauthoryear{{Salpeter}}{{Salpeter}}{1955}]{salpeter55}
{Salpeter} E.~E.,  1955, ApJ, 121, 161

\bibitem[\protect\citeauthoryear{{Scalo}}{{Scalo}}{1986}]{scalo86}
{Scalo} J.~M.,  1986, Fundamentals of Cosmic Physics, 11, 1

\bibitem[\protect\citeauthoryear{{Scholz} \& {Jayawardhana}}{{Scholz} \&
  {Jayawardhana}}{2008}]{scholz08a}
{Scholz} A.,  {Jayawardhana} R.,  2008, ApJL, 672, L49

\bibitem[\protect\citeauthoryear{{Scholz}, {Jayawardhana}, {Muzic}, {Geers},
  {Tamura} \& {Tanaka}}{{Scholz} et~al.}{2012}]{scholz12c}
{Scholz} A.,  {Jayawardhana} R.,  {Muzic} K.,  {Geers} V.,  {Tamura} M.,
  {Tanaka} I.,  2012, ApJ, 756, 24

\bibitem[\protect\citeauthoryear{{Sherry}, {Walter} \& {Wolk}}{{Sherry}
  et~al.}{2004}]{sherry04}
{Sherry} W.~H.,  {Walter} F.~M.,    {Wolk} S.~J.,  2004, AJ, 128, 2316

\bibitem[\protect\citeauthoryear{{Silk}}{{Silk}}{1977}]{silk77a}
{Silk} J.,  1977, ApJ, 214, 152

\bibitem[\protect\citeauthoryear{{Skrutskie}, {Cutri}, {Stiening}, {Weinberg},
  {Schneider}, {Carpenter} \& {25 co-authors}}{{Skrutskie}
  et~al.}{2006}]{skrutskie06}
{Skrutskie} M.~F., et al.\ 2006, AJ, 131, 1163

\bibitem[\protect\citeauthoryear{{Slesnick}, {Carpenter} \&
  {Hillenbrand}}{{Slesnick} et~al.}{2006}]{slesnick06}
{Slesnick} C.~L.,  {Carpenter} J.~M.,    {Hillenbrand} L.~A.,  2006, AJ, 131,
  3016

\bibitem[\protect\citeauthoryear{{Slesnick}, {Hillenbrand} \&
  {Carpenter}}{{Slesnick} et~al.}{2008}]{slesnick08}
{Slesnick} C.~L.,  {Hillenbrand} L.~A.,    {Carpenter} J.~M.,  2008, ApJ, 688,
  377

\bibitem[\protect\citeauthoryear{{Song}, {Zuckerman} \& {Bessell}}{{Song}
  et~al.}{2012}]{song12}
{Song} I.,  {Zuckerman} B.,    {Bessell} M.~S.,  2012, AJ, 144, 8

\bibitem[\protect\citeauthoryear{{Spezzi}, {Alves de Oliveira}, {Moraux},
  {Bouvier}, {Winston}, {Hudelot}, {Bouy} \& {Cuillandre}}{{Spezzi}
  et~al.}{2012}]{spezzi12b}
{Spezzi} L.,  {Alves de Oliveira} C.,  {Moraux} E.,  {Bouvier} J.,  {Winston}
  E.,  {Hudelot} P.,  {Bouy} H.,    {Cuillandre} J.-C.,  2012, A\&A, 545, A105

\bibitem[\protect\citeauthoryear{{Stauffer}, {Schultz} \&
  {Kirkpatrick}}{{Stauffer} et~al.}{1998}]{stauffer98}
{Stauffer} J.~R.,  {Schultz} G.,    {Kirkpatrick} J.~D.,  1998, ApJL, 499, 219

\bibitem[\protect\citeauthoryear{{Walter}, {Vrba}, {Mathieu}, {Brown} \&
  {Myers}}{{Walter} et~al.}{1994}]{walter94}
{Walter} F.~M.,  {Vrba} F.~J.,  {Mathieu} R.~D.,  {Brown} A.,    {Myers} P.~C.,
   1994, AJ, 107, 692

\bibitem[\protect\citeauthoryear{{Whitworth} \& {Stamatellos}}{{Whitworth} \&
  {Stamatellos}}{2006}]{whitworth06}
{Whitworth} A.~P.,  {Stamatellos} D.,  2006, A\&A, 458, 817

\bibitem[\protect\citeauthoryear{{Zapatero Osorio}, {B{\' e}jar},
  {Mart{\'{\i}}n}, {Rebolo}, {Barrado y Navascu{\' e}s}, {Bailer-Jones} \&
  {Mundt}}{{Zapatero Osorio} et~al.}{2000}]{zapatero00}
{Zapatero Osorio} M.~R.,  {B{\' e}jar} V.~J.~S.,  {Mart{\'{\i}}n} E.~L.,
  {Rebolo} R.,  {Barrado y Navascu{\' e}s} D.,  {Bailer-Jones} C.~A.~L.,
  {Mundt} R.,  2000, Science, 290, 103

\end{thebibliography}

\bsp

\label{lastpage}

\end{document}